\documentclass[conference]{IEEEtran}
\IEEEoverridecommandlockouts

\usepackage{cite}
\usepackage{amsmath,amssymb,amsfonts}
\usepackage{algorithmic}
\usepackage{algorithm}
\usepackage{graphicx}
\usepackage{textcomp}
\usepackage{xcolor}
\usepackage{booktabs}
\usepackage{multirow}
\usepackage{subcaption}
\usepackage{mathtools}
\usepackage{bm}

\def\BibTeX{{\rm B\kern-.05em{\sc i\kern-.025em b}\kern-.08em
    T\kern-.1667em\lower.7ex\hbox{E}\kern-.125emX}}

\begin{document}

\title{JCAS-MARL: Joint Communication and Sensing UAV Networks via Resource-Constrained Multi-Agent Reinforcement Learning}

\author{
  \IEEEauthorblockN{Islam Guven, Mehmet Parlak}
  \\
  \IEEEauthorblockA{
    \textit{ICTEAM, Université catholique de Louvain}\\
    Ottignies-Louvain-la-Neuve, Belgium \\
    islam.guven@uclouvain.be
  }
}

\maketitle

\begin{abstract}
Multi-UAV networks are increasingly deployed for large-scale inspection and monitoring missions, where operational performance depends on the coordination of sensing reliability, communication quality, and energy constraints. In particular, the rapid increase in overflowing waste bins and illegal dumping sites has created a need for efficient detection of waste hotspots. In this work, we introduce JCAS-MARL, a resource-aware multi-agent reinforcement learning (MARL) framework for joint communication and sensing (JCAS)-enabled UAV networks. Within this framework, multiple UAVs operate in a shared environment where each agent jointly controls its trajectory and the resource allocation of an OFDM waveform used simultaneously for sensing and communication. Battery consumption, charging behavior, and associated CO$_2$ emissions are incorporated into the system state to model realistic operational constraints. Information sharing occurs over a dynamic communication graph determined by UAV positions and wireless channel conditions. Waste hotspot detection requires consensus among multiple UAVs to improve reliability. Using this environment, we investigate how MARL policies exploit the sensing-communication-energy trade-off in JCAS-enabled UAV networks. Simulation results demonstrate that adaptive pilot-density control learned by the agents can outperform static configurations, particularly in scenarios where sensing accuracy and communication connectivity vary across the environment.\end{abstract}


\section{Introduction}

Unmanned aerial vehicles (UAVs) equipped with 6G sub-terahertz (sub-THz) joint communication and sensing (JCAS) capabilities offer an approach for autonomous navigation, mission-critical data collection, and real-time environmental monitoring applications such as precision agriculture \cite{precision}. UAVs are increasingly used for environmental monitoring tasks including locating overflowing waste bins, detecting illegal dumping sites, and surveying hazardous hotspot regions \cite{waharte2010, goodrich2008, erdelj2017help, martinez2020applications}. Their high mobility, flexible deployment, and low operational cost make them well suited for rapid inspection of large areas. However, limited onboard battery capacity, intermittent connectivity, and variability in sensing performance constrain the effectiveness of UAV operations. These challenges make coordinated multi-UAV exploration and information sharing a dynamic multi-objective optimization problem, where sensing performance, communication reliability, and energy efficiency must be jointly considered.

These operational trade-offs are fundamentally influenced by the underlying communication and sensing mechanisms. Recent developments in JCAS systems provide a physical-layer framework for capturing the interaction between sensing performance and communication capability. Communication-centric JCAS architectures reuse a single OFDM waveform for both radar-style sensing and data communication \cite{zhou_2023_integrated, chen_2020_joint}. Pilot symbols allow echo estimation and improve sensing signal-to-noise ratio (SNR), while data symbols determine throughput. Adjusting the pilot density therefore controls the balance between communication and sensing. Increasing pilot allocation strengthens sensing but reduces communication capacity. Studies on energy-efficient UAV communication further show that such physical-layer choices interact with sustainability constraints \cite{ranjha_2025_consumer}. The literature on environmental UAVs also emphasizes the operational importance of reliable sensing for waste management \cite{gowda2018iotwaste, hodgson2018drones}. 

Multi-agent reinforcement learning (MARL) provides an approach for coordinating UAV teams under these coupled physical constraints \cite{zhang2021multiagentsurvey, lowe2017multiagent}. However, existing MARL environments typically abstract away JCAS effects, assume deterministic sensing, or ignore time-varying communication graphs. Furthermore, these models do not allow control over waveform-level decisions. To address this gap, we propose a sustainability-aware MARL framework in which each UAV controls its motion and its OFDM pilot density. This joint action space enables the policy to modulate sensing and communication quality as operational conditions evolve. Battery usage, renewable-aware charging, and CO$_2$ emissions are integrated into the environment in order to allow carbon-aware decision making during multi-UAV waste hotspot detection. The contributions of this paper are as follows:

\begin{itemize}
    \item We develop a partially-observable MARL environment that optimizes grid-based UAV mobility with an OFDM-based sensing and communication layer, including energy usage, CO$_2$ emissions, and knowledge propagation.
    \item We propose a joint action representation in which each agent controls both its motion and pilot density, allowing adaptive management of sensing quality and communication connectivity under partial observability.
    \item We experimentally demonstrate how PPO policies exploit the waveform capacity across varying fleet sizes and hotspot densities, showing gains in success rate, mission time, and communication efficiency over a constant pilot-density baseline.
\end{itemize}

The remainder of this paper is structured as follows. Section~\ref{sec:system} presents the system model, including the grid layout, the OFDM-based joint sensing-and-communication formulation, and the energy and sustainability framework. Section~\ref{sec:marl} describes the MARL formulation, action and observation spaces, reward design, and the multi-hop knowledge propagation mechanism. Section~\ref{sec:results} reports the experimental results, including sensing–communication trade-offs and sustainability outcomes. Section~\ref{sec:conclusion} concludes the paper and discusses possible extensions.

\begin{figure}[b!]
    \centering
    \includegraphics[width=0.9 \linewidth]{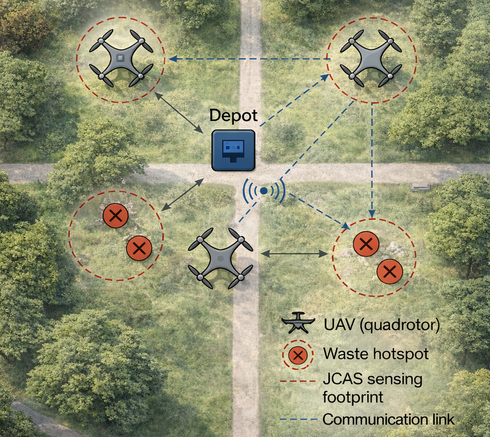}
    \caption{Illustration of the UAV waste-hotspot localization mission, in which a team of UAVs is deployed from a central depot that acts as a base station and patrols the designated grid area.}
    \label{fig:waste_sensing_diagram}
\end{figure}

\section{System Model}
\label{sec:system}
We consider a 2D grid $G = (V, E)$ with square cell size $d$ meters, where $V$ denotes the set of grid cells and $E$ represents the edges between neighboring cells. The grid contains $|V| = W \times H$ cells, with $W$ and $H$ denoting its width and height. A set of $N$ UAVs, $\mathcal{U}=\{1,\dots,N\}$ navigates by moving to one of the 4-neighbors of its current cell or staying. Depots $\mathcal{D}\subset V$ provide charging with electricity under a carbon intensity $c_t$ (kgCO$_2$/kWh) \cite{farras2019carbonaware}.

Each mission starts with a random placement of waste hotspots, $\mathcal{T}^{\text{w}}$. A hotspot is detected when multiple UAVs achieve sufficiently strong JCAS sensing SNR. A schematic overview of the sensing task is shown in Fig.~\ref{fig:waste_sensing_diagram}.

\subsection{JCAS Sensing and Communication Model}
\label{subsec:jcas}

Each UAV transmits a single OFDM waveform that is used both for sensing of waste hotspots and for maintaining a communication link. The OFDM configuration is fixed throughout the mission, the only JCAS-related degree of freedom exposed to the policy is the pilot density $\rho_i^{\text{pilot}}(t)$, which controls how much of the time–frequency grid is reserved for pilots instead of data transmission. The remaining fraction
\[
\rho_i^{\text{comm}}(t) = 1 - \rho_i^{\text{pilot}}(t)
\]
is the communication load. This scalar control ties together sensing reliability and communication quality.

Table~\ref{tab:jcas_params} summarizes the main physical-layer parameters used by the JCAS model and the scale at which they operate. These values are kept constant during training; all adaptation comes from the learned pilot-density decisions and the resulting UAV trajectories.

\begin{table}[b!]
    \centering
    \caption{JCAS parameters.}
    \label{tab:jcas_params}
    \begin{tabular}{lll}
        \toprule
        Symbol & Value & Role \\
        \midrule
        $f_c$ & $5.8~\text{GHz}$ & Carrier frequency of OFDM waveform. \\
        $B$ & $100~\text{MHz}$ & Total bandwidth; sets radar resolution scale. \\
        $P_t$ & $20~\text{dBm}$ & Transmit power per UAV. \\
        $G_t, G_r$ & $2~\text{dBi}$ & TX/RX antenna gains. \\
        $\sigma_j$ & $0~\text{dBsm}$ & Effective radar cross section of hotspot $j$. \\
        $P_n$ & $-90~\text{dBm}$ & Sensing noise floor. \\
        $G_{\text{proc}}$ & $8~\text{dB}$ & OFDM processing gain (coherent combining). \\
        $d$ & $50~\text{m}$ & Physical size of one grid cell. \\
        $n$ & $2.0$ & Path-loss exponent for communication SNR. \\
        $\rho_i^{\text{pilot}}$ & $[0.01,0.30]$ & Pilot density chosen by UAV $i$. \\
        $\alpha_{\text{JCAS}}$ & $1.0~\text{dB}$ & Sensing SNR penalty per unit comm load. \\
        $\gamma_{\text{det}}$ & derived & Effective detection threshold (dB). \\
        $\kappa$ & $0.25$ & Slope of logistic detection curve. \\
        \bottomrule
    \end{tabular}
\end{table}

We adopt a monostatic OFDM radar model. UAV $i$ uses a waveform centered at carrier frequency $f_c$ with 
bandwidth $B$. The received echo power from hotspot $j$ is
\begin{equation}
P_{r,ij}(t) = 
\frac{P_t G_t G_r \lambda^2 \sigma_j}{(4\pi)^3 R_{ij}(t)^4},
\end{equation}
where $\lambda = c/f_c$ and $R_{ij}(t)$ is the UAV--hotspot distance computed from the grid using cell size $d$.

The sensing SNR in dB is then
\begin{equation}
\text{SNR}^{\text{sens}}_{ij}(t)
= 10\log_{10}\!\left(\frac{P_{r,ij}(t)}{P_n}\right)
+ G_{\text{proc}}
- \Delta_{\text{res}}(R_{ij}(t)),
\end{equation}
where $\Delta_{\text{res}}(\cdot)$ is a fixed range-resolution penalty.

Communication load reduces sensing quality by lowering pilot allocation. The effective sensing SNR is
\begin{equation}
\widetilde{\text{SNR}}^{\text{sens}}_{ij}(t)
= \text{SNR}^{\text{sens}}_{ij}(t)
  - \alpha_{\text{JCAS}}\, \rho_i^{\text{comm}}(t).
\end{equation}

The detection margin is defined as
\begin{equation}
m_{ij}(t) = \widetilde{\text{SNR}}^{\text{sens}}_{ij}(t) - \gamma_{\text{det}},
\end{equation}
and mapped to a detection probability via
\begin{equation}
p_{ij}(t) =
\frac{1}{1 + \exp\!\left(-\kappa\, m_{ij}(t)\right)}.
\end{equation}

A hotspot $j$ is considered detected at time $t$ if at least $\theta_{\text{detect}}$ UAVs provide a positive detection:
\begin{equation}
\sum_{i=1}^N \mathbb{1}[\text{detect}_{ij}(t)] \ge \theta_{\text{detect}}.
\end{equation}

Communication uses the same OFDM waveform, with a log-distance path-loss model and noise floor returning pairwise $\text{SNR}^{\text{comm}}_{i, j}(t)$ between each UAV $i$ and $j$. The communication load $\rho_i^{\text{comm}}(t)$ determines how much of the resource grid supports data transfer. These SNR-derived connectivity metrics define the communication graph 
through which detections propagate.

Overall, the MARL policy learns to jointly adjust $\rho_i^{\text{pilot}}(t)$ and UAV trajectories so that sensing confidence, communication reliability, and multi-UAV knowledge propagation are balanced across multiple hotspots. More details on partial observability and knowledge propagation are given in Sec.~\ref{sec:propagation}.

\subsection{Energy and Carbon Model}
Each agent $i$ holds battery capacity $b_i(t)\in[0,B_{\max}]$. Per step,
\begin{align}
e_i(t) &= e_{\text{move}}(a_i(t)) + e_{\text{sense}}(i,t) + e_{\text{comm}}(i,t), \\
b_i(t{+}1) &= \min\{B_{\max},\, b_i(t) - e_i(t) + \chi_i(t)\,r_{\text{ch}}\}, 
\end{align}
where $e_{\text{move}}$ is the propulsion cost (assumed to be constant per move), $e_{\text{sense}}$ depends on pilot density and sensing activity, $e_{\text{comm}}$ accounts for communication overhead, $\chi_i(t)=1$ if at a depot, and $r_{\text{ch}}$ is the charging rate (kWh/step). Charging energy splits into renewable energy share $\rho_{\text{RE}}$ and grid energy:

\begin{equation}
e^{\text{grid}}_i(t) = (1-\rho_{\text{RE}})\,\chi_i(t)\,r_{\text{ch}},\quad
\text{CO}_2(t) = \sum_i e^{\text{grid}}_i(t)\, c_t .
\end{equation}

\subsection{Task Completion}
A hotspot $j$ is detected at $t_j^{\text{det}}$ when the JCAS multi-sensor condition above is met and informed when all agents have the hotspot in their binary knowledge vectors (Sec.~\ref{sec:propagation}). When all hotspots are informed, the mission is completed.

\section{MARL Formulation}\label{sec:marl}
We define a Dec-POMDP $\langle \mathcal{N},\mathcal{S},\{\mathcal{A}_i\},P,R,\{\Omega_i\},O,\gamma\rangle$ \cite{zhang2021multiagentsurvey, lowe2017multiagent}. Each agent controls both motion and JCAS waveform configuration.

\subsection{Action Space}
Each agent $i$ selects a continuous 2D action:
\begin{equation}
\mathbf{a}_i(t) = (u_i^{\text{dir}}(t), u_i^{\text{pilot}}(t)) \in [-1,1]^2.
\end{equation}
The first component $u_i^{\text{dir}}$ is mapped to one of five motions $\{\text{up},\text{down},\text{left},\text{right},\text{stay}\}$. The second component $u_i^{\text{pilot}}$ is mapped to a pilot density
\begin{equation}
\rho_i^{\text{pilot}}(t) = \rho_{\min} + \frac{u_i^{\text{pilot}}(t)+1}{2}\,(\rho_{\max}-\rho_{\min}),
\end{equation}
clipped to $[\rho_{\min},\rho_{\max}]$, with $\rho_i^{\text{comm}}(t)=1-\rho_i^{\text{pilot}}(t)$. Therefore, the MARL policy must co-design UAV trajectories and JCAS resource allocation.

\subsection{Reward Design with JCAS and Sustainability Signals}\label{sec:marl_reward}
We combine sparse task outcomes with dense JCAS and sustainability shaping:
\begin{align}
R_t &= 
\underbrace{\sum_{j\in \mathcal{T}^{\text{w}}}\!\!\Big(
    \alpha_{\text{det}}\mathbb{1}[t{=}t_j^{\text{det}}]
  + \alpha_{\text{inf}}\mathbb{1}[\text{informed}_j(t)]
\Big)}_{\text{waste detection \& informing}} \nonumber\\
&\quad 
+ \underbrace{\alpha_{\text{comp}}\,\mathbb{1}[\text{all hotspots informed at } t]}_{\text{completion reward}}
\nonumber\\
&\quad 
+ \underbrace{\alpha_{\text{cov}} \cdot \text{coverage\_gain}(t)}_{\text{coverage reward}} \nonumber\\
&\quad 
- \underbrace{\eta_e \sum_i e_i(t)}_{\text{energy penalty}} 
- \underbrace{\eta_c \sum_i e_i^{\text{grid}}(t)\, c_t}_{\text{carbon penalty}} \nonumber\\
&\quad 
- \underbrace{\xi_{\text{rev}} \cdot \text{revisit\_count}(t)
  + \xi_{\text{trunc}} \,\mathbb{1}[t{=}T_{\max}]}_{\text{efficiency penalties}} \nonumber\\
&\quad 
+ \underbrace{\zeta_{\text{comm}} \cdot \overline{\eta}(t)}_{\text{communication throughput reward}}
\nonumber\\
&\quad 
+ \underbrace{\zeta_{\text{spread}} \cdot \text{knowledge\_spread}(t)}_{\text{propagation bonus}}.
\end{align}

where $\overline{\eta}(t)$ is the average normalized communication throughput across agents at time $t$, and $\text{knowledge\_spread}(t)$ measures the fraction of agents that know each detected hotspot. We use potential-based terms for distance-to-goal (nearest undetected hotspot) and for carbon-aware charging (encouraging charging at depots with high $\rho_{\text{RE}}$ or low $c_t$) to preserve optimality \cite{ng1999policy}.

\subsection{Observations}
Each agent receives a local observation $\mathbf{o}_i(t)$ including:
\begin{itemize}
    \item \textbf{Self:} normalized position on the grid, battery SoC $b_i(t)/B_{\max}$, current pilot density $\rho_i^{\text{pilot}}(t)$, and a coarse estimate of its own communication SNR or throughput.
    \item \textbf{Targets:} for each waste hotspot $j$: relative grid position, detection status, and agent knowledge bit $k_{ij}(t)$.
    \item \textbf{Other UAVs/Comm:} relative positions of neighbors within communication radius, and an aggregated measure of JCAS connectivity.
    \item \textbf{Sustainability:} current carbon intensity $c_t$ and distance to nearest depot.
    \item \textbf{Global progress:} fractions of detected/informed hotspots, and normalized time index.
\end{itemize}

\subsection{Knowledge Propagation}
\label{sec:propagation}
After JCAS sensing, each agent updates its binary knowledge vector $\mathbf{k}_i(t)$ based on local detections, then runs a multi-iteration consensus step over the communication graph. $\mathcal{N}_i(t)$ is the communication neighbors of agent $i$ at time $t$ and $\mathbf{k}_i^{(0)}(t)$ the post-sensing vector. For $\ell=1,\dots,N$,
\begin{equation}
\mathbf{k}_i^{(\ell)}(t) \leftarrow \mathbf{k}_i^{(\ell-1)}(t) \bigvee_{j\in \mathcal{N}_i(t)} \mathbf{k}_j^{(\ell-1)}(t),
\end{equation}

where $\bigvee$ denotes the elementwise logical OR. Therefore, each connected component of the communication graph becomes aware of the detected hotspots in a single environment step \cite{olfati2007consensus}.

\subsection{Training Algorithm}
We use centralized training with decentralized execution (CTDE) \cite{kraemer2016multi}. Policies are shared across agents and trained with Synchronous PPO with GAE \cite{schulman2017proximal, schulman2015high}, collecting batches synchronously from all actors. Policies use 3-layer MLP architectures (512-256-128 units). We implement the MARL training pipeline using Ray RLlib \cite{liang2018rllibabstractionsdistributedreinforcement}.

\section{Results}
\label{sec:results}
\subsection{Environment setup}

\begin{table}[b!]
\centering
\caption{Experiment Configuration for Sustainability-Aware JCAS–MARL Training and Evaluation}
\label{tab:exp_config}
\begin{tabular}{llc}
\toprule
\textbf{Category} & \textbf{Parameter} & \textbf{Value} \\
\midrule
\multicolumn{3}{l}{\textbf{Environment Configuration}} \\
\midrule
Grid size & $W \times H$ & $12 \times 12$ cells \\
Cell size & $d$ & $50$ m \\
Number of UAVs & $N$ & $\{5, 10, 15, 20\}$ \\
Number of hotspots & $n_{\text{targets}}$ & $\{3, 5, 7, 9, 11\}$ \\
Max steps per episode & $T_{\max}$ & $100$ \\
Detection threshold & $\theta_{\text{detect}}$ & $3$ UAVs \\
\midrule

\multicolumn{3}{l}{\textbf{Energy and Sustainability Model}} \\
\midrule
Battery capacity & $B_{\max}$ & $0.20$ kWh \\
Return-to-base threshold &  & $0.04$ kWh \\
Charge rate & $r_{\text{ch}}$ & $0.30$ kW \\
Movement cost & $e_{\text{move}}$ & $8\!\times\!10^{-4}$ kWh \\
Sensing energy & $e_{\text{sense}}$ & $2\!\times\!10^{-4}$ kWh \\
Communication energy & $e_{\text{comm}}$ & $5\!\times\!10^{-5}$ kWh \\
Renewable energy share & $\rho_{\text{RE}}$ & $0.1$ \\
Grid CO$_2$ intensity & $c_t$ & $[0.25-0.40]$ kg/kWh \\
\midrule
\multicolumn{3}{l}{\textbf{Reward Weights}} \\
\midrule
Correct detection reward &  & $7.0$ \\
Inform reward &  & $4.0$ \\
Completion reward &  & $10.0$ \\
Coverage reward &  & $0.5$ \\
Revisit penalty &  & $-0.01$ \\
Truncation penalty &  & $-0.4$ \\
Energy penalty coefficient &  & $0.2$ \\
Carbon penalty coefficient &  & $0.1$ \\
Communication throughput reward &  & $0.5$ \\
Knowledge spread reward &  & $0.1$ \\
\midrule
\multicolumn{3}{l}{\textbf{MARL / PPO Configuration}} \\
\midrule
Algorithm &  & PPO (CTDE) \\
Discount factor & $\gamma$ & $0.95$ \\
Learning rate & $\alpha$ & $3\times10^{-4}$ \\
Batch size &  & $4096$ \\
Training steps &  & $2.5\times 10^5$ \\
Episodes per evaluation & $M$ & $100$ \\

\bottomrule
\end{tabular}
\end{table}

We consider a $12{\times}12$ grid with cell size $d=50$m and a mission horizon of $T_{\max}=100$, which corresponds to the operational endurance implied by a 0.20kWh battery per UAV. We evaluate fleets of $N\in{5,10,15,20}$ UAVs and hotspot configurations with $n_{\text{targets}}\in{3, 5,7,9,11}$ waste locations generated randomly for each episode. 

The carbon intensity of the charging station (depot) $c_t$ is sampled uniformly in the range
$[0.25, 0.40]$ kg/kWh at each step, mimicking a time-varying grid mix. 

The joint communication-and-sensing layer uses a single OFDM waveform at $5.8$GHz with $100$MHz bandwidth with parameters defined in Sec.~\ref{sec:system}. Each UAV controls only its pilot density, which is bounded within $[0.01,0.30]$ and determines the sensing–communication resource split. A hotspot is confirmed when at least $\theta_{\text{detect}}=3$ UAVs detect it in the same step, and a mission is successful only if all hotspots are detected and informed to all UAVs. Additional parameters are included in Table \ref{tab:exp_config}. 

\subsection{Training Costs}

Table~\ref{tab:train_iterations} summarizes the PPO training costs for different fleet sizes on our Intel Xeon Gold 6444Y CPU with 8 parallel processes. Each evaluation rollout takes on the order of $1$--$2$\,s.

\begin{table}[h!]
    \centering
    \caption{PPO training times in seconds for different fleet sizes.}
    \label{tab:train_iterations}
    \begin{tabular}{lcccc}
        \toprule
        Method & $N{=}5$ & $N{=}10$ & $N{=}15$ & $N{=}20$ \\
        \midrule
        PPO        & 402  & 598 & 1253 & 1751 \\
        \bottomrule
    \end{tabular}
\end{table}

\subsection{PPO Convergence Across Target Densities}

\begin{figure}[b!]
        \centering
        \includegraphics[width=0.9 \linewidth]{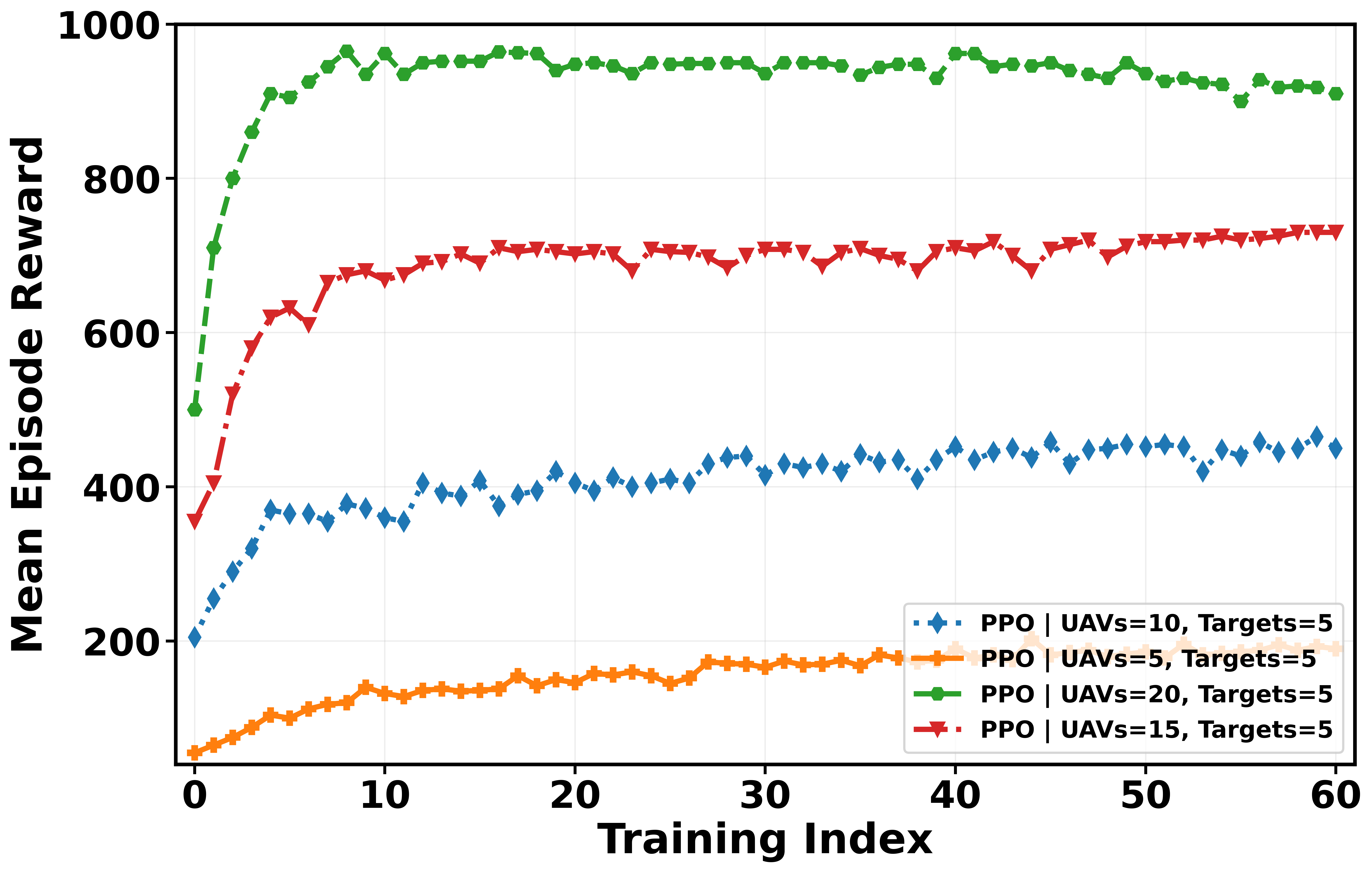}

        \caption{Training evolution of the mean episode return for different numbers of targets.}
    \label{fig:train_returns}
\end{figure}

Fig.~\ref{fig:train_returns} shows the evolution of the mean episode return as a function of PPO training iterations for different numbers of hotspots. One PPO iteration corresponds to sampling one full batch of 4096 environment steps and performing 10 epochs of gradient updates. Across all target sets, the return stabilizes after around 40 iterations, indicating that PPO can reliably learn a joint trajectory and pilot-density policy in this JCAS-enabled environment.

\subsection{Effect of Fleet Size on Detection Performance}

\begin{figure}
    \centering
    \includegraphics[width=0.9\linewidth]{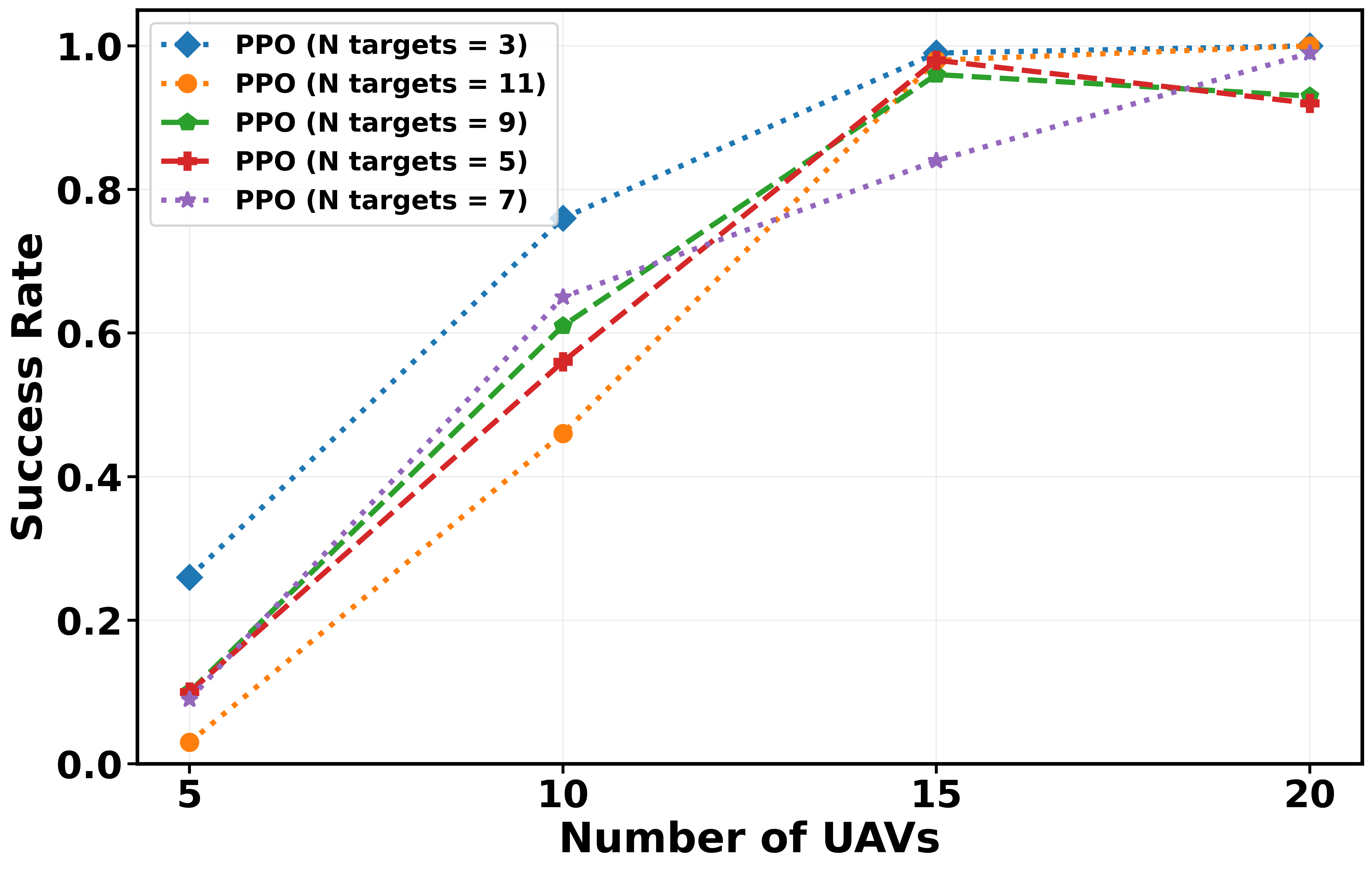}
    \caption{Evaluation success rate as a function of the number of UAVs for $3-11$ targets.}
    \label{fig:success_uavs}
\end{figure}

We next evaluate the learned PPO policies in terms of mission success. 
Fig.~\ref{fig:success_uavs} reports the evaluation success rate as a function of the number of UAVs for $3-11$ targets on the same region.
The curves show that fleets of $10$ UAVs already achieve an average success rate around $0.73$, while $15$ UAVs reach approximately $97\%$ success for all three target densities.

\begin{figure}[t]
    \centering
    \includegraphics[width=0.9\linewidth]{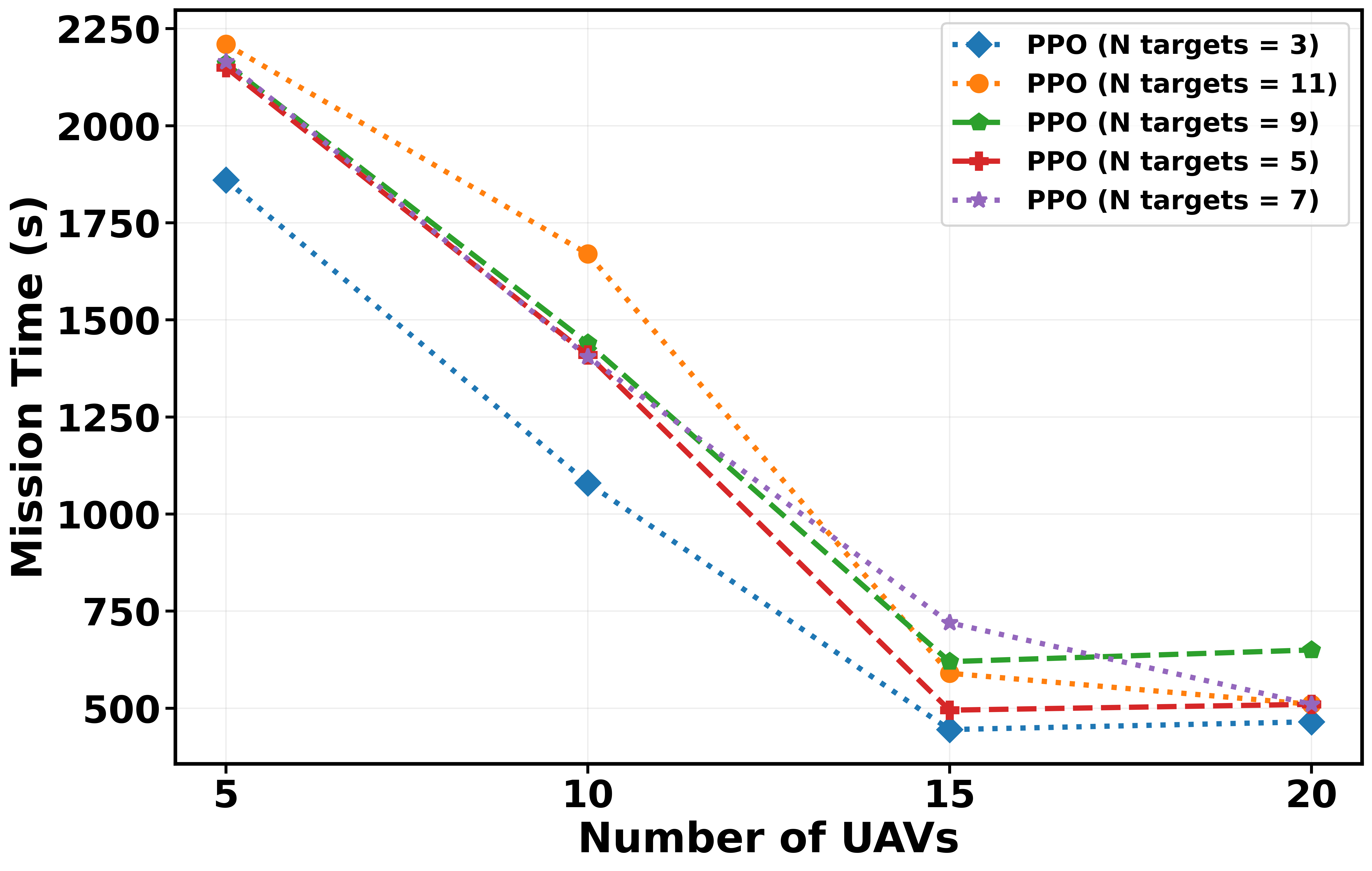}
    \caption{Evaluation mission time as a function of the number of UAVs.}
    \label{fig:mission_uavs}
\end{figure}

Fig.~\ref{fig:mission_uavs} shows the corresponding mission times. 
As the fleet size grows, mission time decreases sharply and then saturates once the team is large enough to cover the region and confirm all hotspots without long waiting times.

\subsection{Energy Consumption and Communication Throughput}

\begin{figure}
    \centering
    \includegraphics[width=0.9\linewidth]{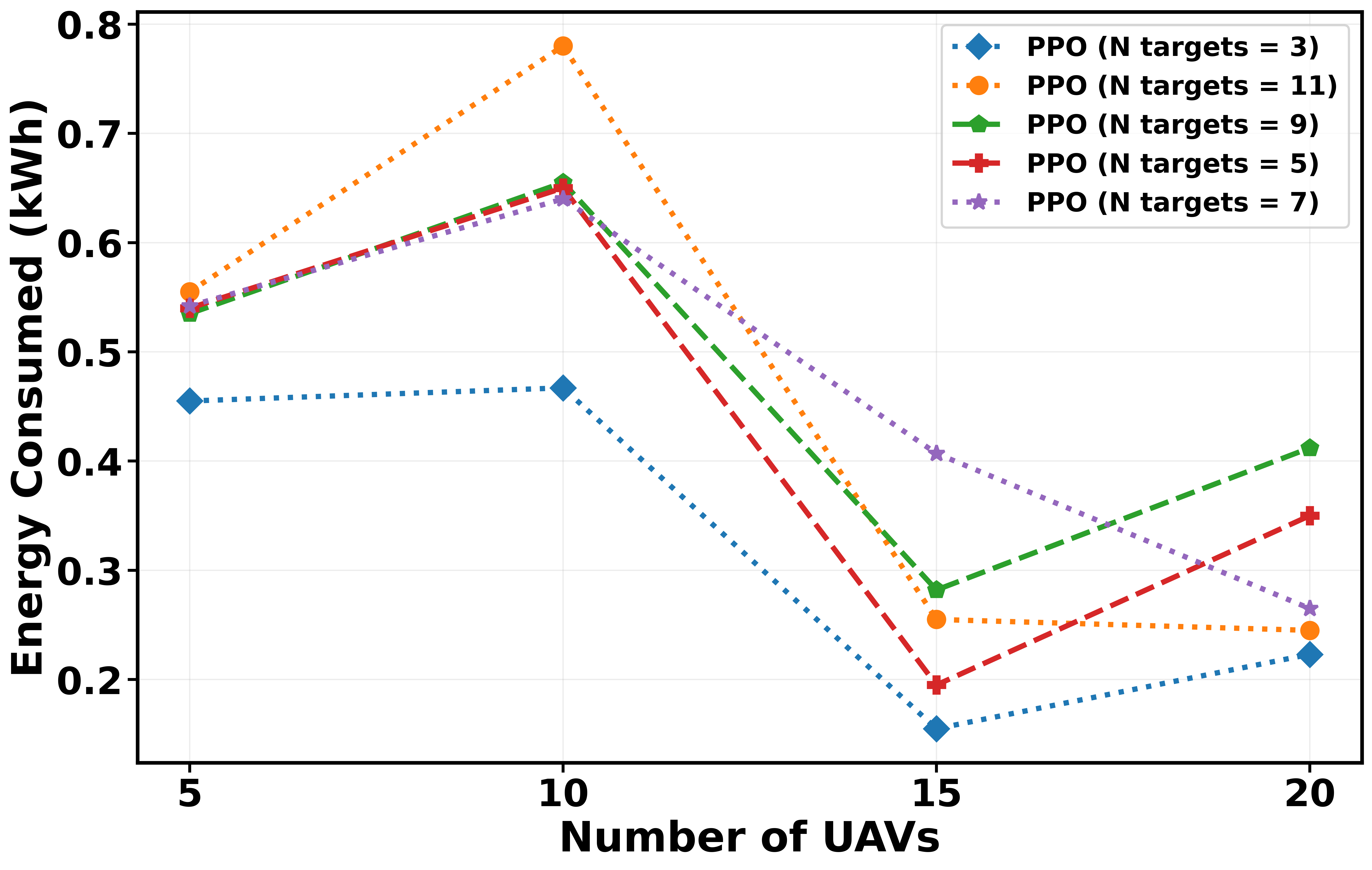}
    \caption{Total energy consumption per mission as a function of the number of UAVs.}
    \label{fig:energy_uavs}
\end{figure}

To quantify sustainability, Fig.~\ref{fig:energy_uavs} reports the total energy consumption per mission versus the number of UAVs. 
Although larger fleets consume more energy in aggregate, the increase is sublinear in $N$ because better coverage reduces redundant motion and repeated hotspot visits.

\begin{figure}
    \centering
    \includegraphics[width=0.9\linewidth]{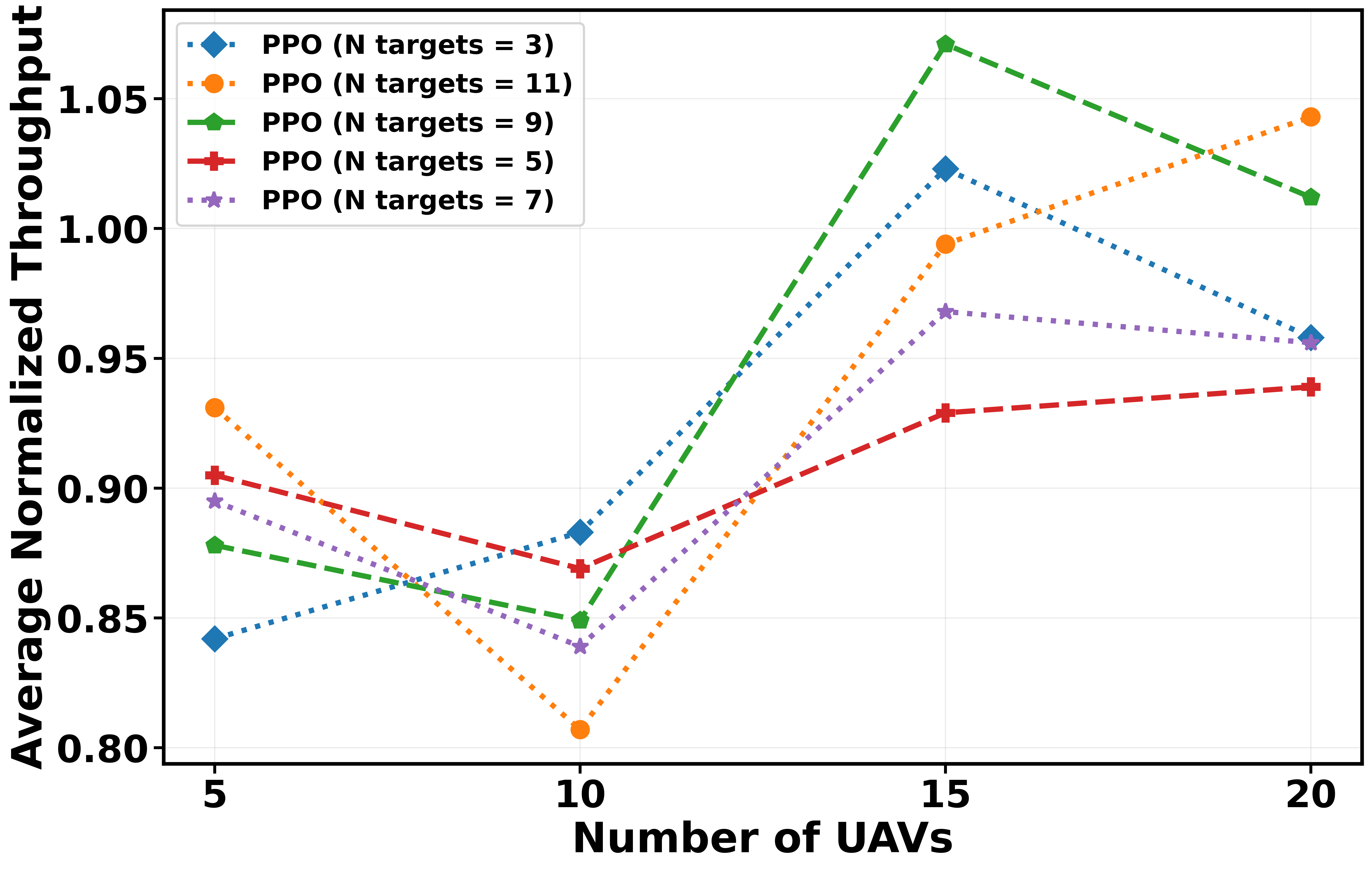}
    \caption{Normalized communication throughput as a function of the number of UAVs.}
    \label{fig:throughput_uavs}
\end{figure}

Finally, Fig.~\ref{fig:throughput_uavs} illustrates the normalized communication throughput obtained from the OFDM-based JCAS model as a function of fleet size. The learned policies keep pilot density high when approaching suspected hotspots and reduce it afterwards, preserving enough throughput to propagate detections through the communication graph even for larger teams. Because throughput is normalized using a fixed reference spectral efficiency of 6 bits/s/Hz, episodes with high communication SNR can return normalized throughput values above 1.

\begin{figure}[b!]
    \centering
    \includegraphics[width=0.9\linewidth]{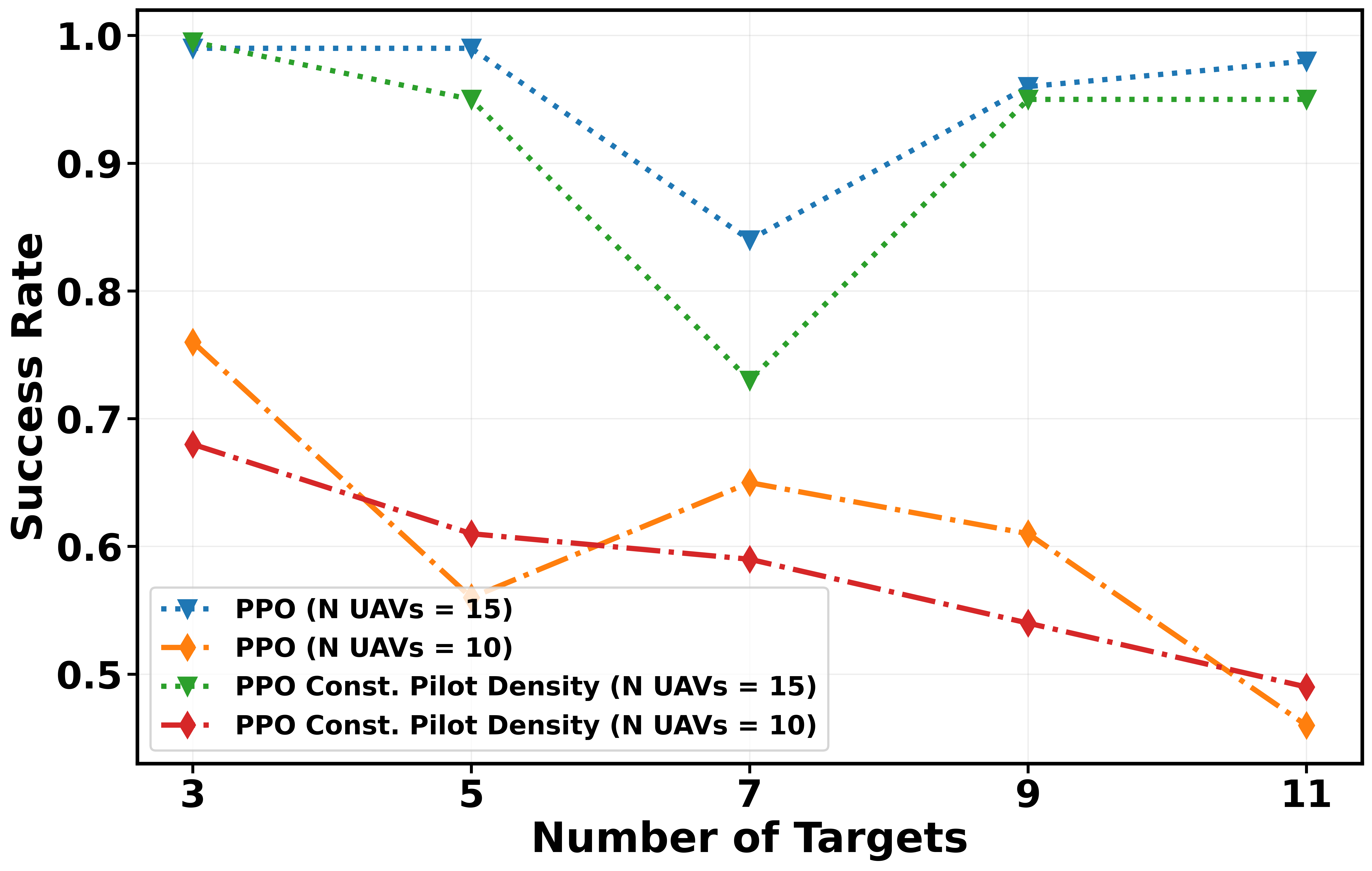}
    \caption{Evaluation success rate as a function of the number of Hotspots.}
    \label{fig:success_targets}
\end{figure}

\subsection{Constant Pilot Density Model}

We also analyze the success rate of a constant pilot density model where pilot density is set at $0.3$ for allowing both communication and sensing without adaptivity. Fig.~\ref{fig:success_targets} shows that with larger fleets, the improvement of success rates reach around 10\% while for $10$ UAVs, both models have fluctuating advantages around 5\%. This shows that adaptivity is not always an advantage, and larger target numbers reduce the need for adaptivity since the mission gets closer to an area coverage mission.

\begin{figure}
    \centering
    \includegraphics[width=0.9\linewidth]{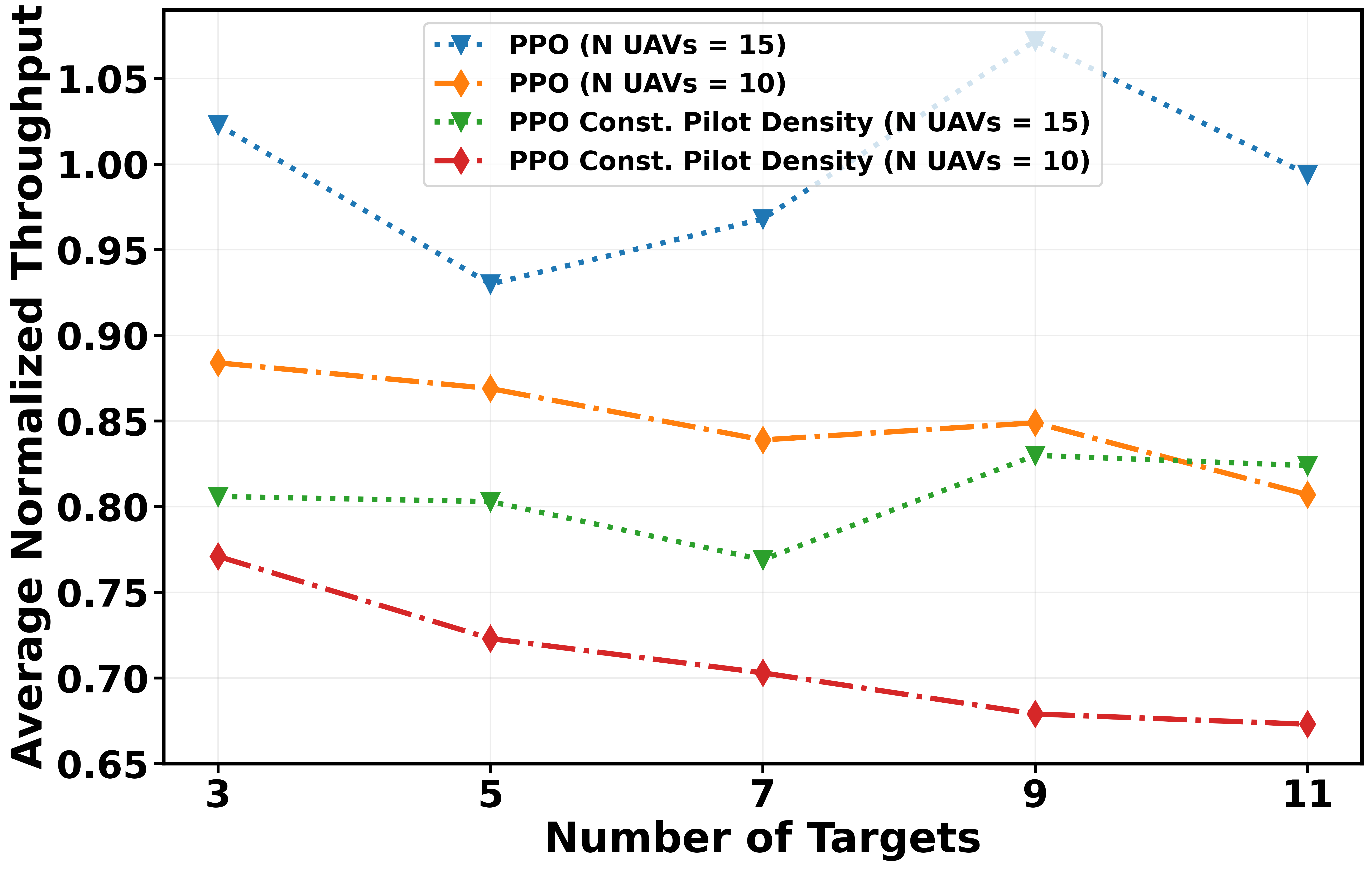}
    \caption{Normalized communication throughput as a function of the number of Hotspots.}
    \label{fig:throughput_targets}
\end{figure}

To further analyze the effect of throughput on pilot density, we compare the average throughput for $10$ and $15$ UAVs for both configurations in Fig.~\ref{fig:throughput_targets}. With a constant pilot rate, UAVs can no longer take advantage of boosting communication power on situations where there is no need for sensing which results in a reduced average throughput.

\section{Conclusion}
\label{sec:conclusion}
We demonstrated a sustainability- and JCAS-aware MARL framework for UAV-based waste hotspot response that integrates battery dynamics, energy consumption, carbon emissions, and joint communication-and-sensing signals into the learning process. A monostatic OFDM radar--communication model replaces range-based abstractions, allowing agents to adapt pilot density to balance sensing reliability and communication throughput.

Simulation results in a $12{\times}12$ grid environment with fleets of $5$--$20$ UAVs and $3$--$11$ hotspots show that PPO agents learn stable policies within about $40$ training iterations, with training times ranging from $402$\,s to $1751$\,s depending on fleet size. The learned policies achieve high mission success rates, reaching approximately $0.73$ with $10$ UAVs and about $97\%$ with $15$ UAVs across target densities, while reducing mission time as fleet size increases. Despite larger teams consuming more total energy, the growth is sublinear due to improved spatial coverage. Furthermore, adaptive pilot-density control maintains reliable communication throughput while enabling multi-sensor confirmation of hotspots, outperforming fixed pilot configurations in several scenarios.

These results demonstrate that integrating sustainability-aware objectives and JCAS resource control within MARL enables coordinated UAV policies that effectively detect waste hotspots while balancing sensing reliability, communication performance, and energy use.

\section*{Acknowledgment}
This work was supported by the Brains for Brussels research and innovation funding program of the Région de Bruxelles-Capitale–Innoviris under Grant RBC/BFB 2023-BFB-2.

\bibliographystyle{IEEEtran}



\end{document}